\documentclass[twocolumn]{aastex63}
\def \colw {0.5 \textwidth}
\usepackage[english]{babel}
\usepackage[utf8x]{inputenc}
\usepackage[T1]{fontenc}
\usepackage{textgreek}
\usepackage{amsmath}
\usepackage{amssymb}
\usepackage{graphicx}
\usepackage{commath}
\usepackage{multirow}
\usepackage{color}
\usepackage{aecompl}
\usepackage{booktabs}
\usepackage[colorinlistoftodos]{todonotes}
\usepackage{tikz}

\def\be{\begin{equation}}
\def\ee{\end{equation}}
\def\ba{\begin{eqnarray}}
\def\ea{\end{eqnarray}}
\def\go{\mathrel{\raise.3ex\hbox{$>$}\mkern-14mu
             \lower0.6ex\hbox{$\sim$}}}
\def\lo{\mathrel{\raise.3ex\hbox{$<$}\mkern-14mu
             \lower0.6ex\hbox{$\sim$}}}

\def\bxi{{\mbox{\boldmath $\xi$}}}
\def\br{{\bf r}}
\def\bC{{\bf C}}
\def\bOmega{{\bf \Omega}}
\def\Oms{\Omega_s}
\def\Omo{\Omega_{\rm orb}}
\def\omi{\omega_\alpha}

\def\red#1 {\textcolor{red}{#1}\ }   
\def\blue#1 {\textcolor{blue}{#1}\ }   

\shorttitle{Jupiter Love Number}
\shortauthors{Dong Lai}
\begin{document}
\title{Jupiter's Dynamical Love Number}

\author[0000-0002-1934-6250]{Dong Lai}

\affil{Cornell Center for Astrophysics and Planetary Science, Department of Astronomy,
Cornell University, Ithaca, NY 14853}


\begin{abstract}
Recent observations by the {\it Juno} spacecraft have revealed that
the tidal Love number $k_2$ of Jupiter is $4\%$ lower than the
hydrostatic value.  We present a simple calculation of the dynamical
Love number of Jupiter that explains the observed ``anomaly''.  The
Love number is usually dominated by the response of the
(rotation-modified) f-modes of the planet. Our method also allows for
efficient computation of high-order dynamical Love numbers. While the
inertial-mode contributions to the Love numbers are negligible, 
a sufficiently strong stratification in a large region of the planet's
interior would induce significant g-mode responses and influence the
measured Love numbers.
\end{abstract}


\section{Introduction}

The {\it Juno} spacecraft recently found an ``anomaly'' in Jupiter's
tidal Love number: the measured $k_2=0.565\pm 0.006$ (Durante et
al.~2020) appears to be smaller than the theoretical hydrostatic value
$k_2^{\rm (hs)} =0.590$ (Wahl et al.~2020) by $4\%$.  This discrepancy
may be explained in terms of dynamical tides, i.e., Jupiter's response
to the finite-frequency tidal forcings from the Galilean moons (Idini \&
Stevenson 2021).  Here we present a simple calculation that explains
this Love number ``anomaly'' quantitatively. Naive expectation would
suggest a $1/(\omega_\alpha^2-\omega^2)$ enhancement (where
$\omega_\alpha$ is the f-mode frequency of the planet) of the tidal
response due to the finite tidal frequency ($\omega$) as compared to
the hydrostatic ($\omega=0$) response.  The key to obtain the correct
answer is to treat the rotational (Coriolis) effect on the modes of a rotating
planet and their tidal responses in a self-consistent way. Our general
method also allows for efficient computation of high-order dynamical Love
numbers $k_{lm}$, as well as the inclusion of the contributions to
$k_{lm}$ from the inertial modes (due to planetary rotation) and g-modes (due to
stable stratification in the planetary interior).

\section{Dynamical Love Number and Normal Modes}

Consider a planet (mass $M$, radius $R$ and spin angular frequency $\Oms$)
orbited by a satellite (mass $M'$) in a circular
orbit with semi-major axis $a$ and orbital frequency $\Omo$. We assume the spin
axis is aligned with the orbital axis. In the frame corotating with the planet,
the $(lm)$-component of the tidal potential produced by $M'$ on the planet is
\be
U(\br,t)
=-A_{lm}\,r^l\,Y_{lm}(\theta,\phi)\, e^{-i\omega t},
\label{eq:potential}
\ee
where $A_{lm}=(GM'/a^{l+1})W_{lm}$ (with $W_{lm}$ a dimensionless constant;
$W_{lm}\neq 0$ when $l+m=$even),
$\br=(r,\theta,\phi)$ specifies the position vector (in
spherical coordinates) measured from the center of the planet,
and
\be
\omega=m(\Omo-\Oms)
\ee
is the tidal forcing frequency. It suffices to consider only $m>0$.
The relevant non-zero tidal components are $(lm)=(2,2),(3,1),(3,3),(4,2),(4,4)$ etc.

The linear response of the planet to the tidal forcing
is specified by the Lagrangian displacement, $\bxi(\br,t)$, of a fluid element
from its unperturbed position.  In the rotating frame of the planet, the
equation of motion takes the form
\be
\frac{\partial^2 \bxi}{\partial t^2}+2\bOmega_s\times
\frac{\partial\bxi}{\partial t}+{\bC}\cdot\bxi=-\nabla U,
\label{eq:eqnmotion2}
\ee
where $\bC$ is a self-adjoint operator (a function of the pressure and
gravity) acting on $\bxi$ (see, e.g., Friedman \& Schutz 1978).
A free mode of frequency $\omega_\alpha$ (in the rotating frame)
with $\bxi_\alpha(\br,t)=\bxi_\alpha(\br)\,e^{-i\omega_\alpha t}\propto
e^{im\phi-i\omega_\alpha t}$ satisfies
\be
-\omi^2\bxi_\alpha-2i\omi\bOmega_s\times\bxi_\alpha+\bC\cdot
\bxi_\alpha=0,
\ee
where $\{\alpha\}$ denotes the mode index, which includes the
azimuthal number $m$. We carry out phase-space mode expansion
(Schenk et al.~2002)
\be
\left[\begin{array}{c}
\bxi\\
{\partial\bxi/\partial t}
\end{array}\right]
=\sum_\alpha c_\alpha(t)
\left[\begin{array}{c}
\bxi_\alpha(\br)\\
-i\omi\bxi_\alpha(\br)
\end{array}\right].
\ee
Using the orthogonality relation 
$\langle\bxi_\alpha,2i\bOmega_s\times\bxi_{\alpha'}\rangle+
(\omega_\alpha+\omega_{\alpha'})\langle\bxi_\alpha,\bxi_{\alpha'}
\rangle=0$ (for $\alpha\neq \alpha'$), where                                                             
$\langle A,B\rangle\equiv\int\!d^3x\,\rho\, (A^\ast\cdot B)$,
we find (Lai \& Wu 2005)
\be
{\dot c}_\alpha+i\omi c_\alpha =
{i Q_{\alpha,lm}\over 2\varepsilon_\alpha}\, A_{lm}\, e^{-i\omega t},
\label{eq:adot}
\ee
where
\ba
&&Q_{\alpha,lm}\equiv\bigl\langle\bxi_\alpha,\nabla (r^lY_{lm})
\bigr\rangle = \int\!d^3x\, r^l\, Y_{lm}\,\delta\rho^\ast_\alpha,
\label{eq:Qdefine}\\
&& \varepsilon_\alpha\equiv
\omi+\langle\bxi_\alpha,i\bOmega_s\times\bxi_{\alpha}\rangle,
\ea
and we have used the normalization $\langle\bxi_\alpha,\bxi_\alpha\rangle                                
=1$. In Eq.~(\ref{eq:Qdefine}), $\delta\rho_\alpha$ is the Eulerian density perturbation
associated with the eigenfunction $\bxi_\alpha$.

Equation (\ref{eq:adot}) has stationary solution
\be
c_\alpha (t)={Q_{\alpha,lm}\over 2\varepsilon_\alpha (\omega_\alpha-\omega)}\, A_{lm}\, e^{-i\omega t}.
\ee
The gravitational perturbation associated with the density perturbation
$\delta\rho(\br,t)=\sum_\alpha c_\alpha(t)\delta\rho_\alpha(\br)$, evaluated at the planet's surface
($r=R$), is
\be
\delta\Phi(\br,t)\Bigr|_{r=R}= - \sum_\alpha
c_\alpha (t)\, {4\pi\over 2l+1}\, {GM Q_{\alpha,lm}\over R}\, Y_{lm}.
\ee
Thus the tidal Love number is
\be
k_{lm}={\delta\Phi\over U}\Bigr|_{r=R}={2\pi \over 2l+1}\sum_\alpha {\bar Q_{\alpha,lm}^2\over
  \bar\varepsilon_\alpha (\bar\omega_\alpha-\bar\omega)}.
\label{eq:klm}\ee
In the above equation, the tidal overlap coefficient $\bar Q_{\alpha,lm}$ and the mode frequencies
$\bar\omega_\alpha$ and $\bar\varepsilon_\alpha$ are in units where
$G=M=R=1$, i.e., $\bar\omega_\alpha=\omega_\alpha/(GM/R^3)^{1/2}$, etc.
Note that for a given $m>0$, the sum in Eq.~(\ref{eq:klm}) includes modes with positive $\omega_\alpha$
and negative $\omega_\alpha$, corresponding to prograde (with respect to the planet's rotation)
and retrograde modes.

\section{F-mode Contribution}


\begin{deluxetable}{ccccc}[!t]
\tablecaption{Oscillation modes of non-rotating polytropic $(n=1)$ planet model\label{tab:table1}}
\tablehead{ & & $\omega_0 $ & $Q_l$ & $C$ }
\startdata
$\Gamma_1=2$ & & & & \\
$l=2$ & f &  0.1227E+01 &  0.5579E+00 & 0.4991E+00\\
      & p1&  0.3462E+01 &  0.2690E-01 & 0.1119E+00\\
$l=3$ & f &  0.1698E+01 &  0.5846E+00 & 0.3321E+00\\
      & p1& 0.3975E+01&  0.4054E-01&  0.8627E-01\\
$l=4$ & f &  0.2037E+01 &  0.5979E+00 & 0.2489E+00\\
      &p1 & 0.4409E+01&  0.4623E-01 & 0.6984E-01\\
$\Gamma_1=2.4$ & & & & \\ $l=2$
&f& 0.1230E+01& 0.5580E+00& 0.4985E+00\\
&g1& 0.4688E+00& -0.1313E-01& 0.1057E+00\\
&g2& 0.3270E+00& 0.3071E-02 & 0.1342E+00\\
&g3& 0.2526E+00& -0.8961E-03 & 0.1464E+00\\
$l=3$ &f& 0.1703E+01& 0.5850E+00& 0.3317E+00\\
&g1& 0.5681E+00& -0.1216E-01& 0.3321E-01\\
&g2& 0.4117E+00& 0.3257E-02& 0.5622E-01\\
&g3& 0.3254E+00& -0.1031E-02& 0.6603E-01\\
$\Gamma_1=2.4$& for & $r\in [0.5,0.7],$& $[0.85,0.93]~~~~$& \\
$l=2$ &f& 0.1228E+01 & 0.5580E+00 & 0.4988E+00\\
&g1& 0.3664E+00 & -0.6986E-02 & 0.9279E-01\\
&g2& 0.2022E+00 & 0.1408E-02 & 0.1207E+00\\
&g3& 0.1565E+00 & -0.1546E-03 & 0.1467E+00\\
$l=3$ &f &0.1701E+01 & 0.5849E+00 & 0.3319E+00\\
&g1& 0.4489E+00& -0.6067E-02& 0.2790E-01\\
&g2& 0.2749E+00& 0.2049E-02& 0.4166E-01\\
&g3& 0.2126E+00& -0.1567E-03& 0.6622E-01\\
\enddata
\tablecomments{$\omega_0$ and $Q_l$ are the mode frequency and tidal
  overlap coefficient (Eq.~\ref{eq:Qdefine}), both in units such that
  $G=M=R=1$, and $C$ is defined in Eq.~(\ref{eq:c}).  The planet's
  density profile is that of $n=1$ polytrope (with the equation of
  state $P\propto \rho^2$). The first model has $\Gamma_1$ (the
  adiabatic index) equal to $\Gamma=1+1/n$, and we list the properties
  for the f-mode and the first radial-order p-mode. The second model
  has $\Gamma_1=2.4$ throughout the planet, and the third model has
  $\Gamma_1=2.4$ only in two regions ($r/R\in
  [0.5,0.7],\,[0.85,0.93]$) and $\Gamma_1=\Gamma$ otherwise
  (the transition width is $0.025R$; see Eq.~\ref{eq:gam1}),
  and we list the properties for the f-mode
  and the first three radial-order g-modes. Note that when $|Q_l|\ll 1$, the quoted $Q_l$ values
  are only accurate in 2-3 significant figures.}
\end{deluxetable}

In most situations, the sum in Eq.~(\ref{eq:klm}) is dominated by f-modes since they have the largest
tidal overlap $Q_{\alpha,lm}$. For planetary rotation rate $\Oms$ much less than the breakup rate $(GM/R^3)^{1/2}$,
(e.g., $\bar\Omega_s=0.288$ for Jupiter),
the effect of rotation on the modes can be treated perturbatively
(e.g. Unno et al.~1989). Let $\omega_0$ ($>0$) be the mode
frequency of a non-rotating planet, then for a given $m>0$, the sum in Eq.~(\ref{eq:klm}) includes
\ba
&&\varepsilon_\alpha\simeq \pm \omega_0, \label{eq:eps}\\
&&\omega_\alpha=\pm \omega_0 -m C \Oms,\label{eq:omeg}
\ea
with
\ba
mC &\equiv & \int\!d^3x\,\rho\,\bxi^\ast_{\alpha,0}\cdot (i{\bf\hat z}\times\bxi_{\alpha,0})\nonumber\\
&=& m\int_0^R\!\!dr\,\rho r^2(2\xi_r\xi_\perp+\xi_\perp^2), \label{eq:c}
\ea
where $\bxi_{\alpha,0}=\left[\xi_r(r){\bf\hat r}+\xi_\perp(r) r\nabla\!_\perp\right]Y_{lm}$
is the mode eigenvector of a non-rotating planet. To a good
approximation, we can also set $Q_{\alpha,lm}$ to be the non-rotating value, i.e,
\be
Q_{\alpha,lm}\simeq Q_l.
\ee
Thus Eq.~(\ref{eq:klm}) reduces to 
\be
k_{lm}\simeq \left({4\pi\over 2l+1}\right) {\bar Q_l^2\over \bar\omega_0^2-(mC\bar\Omega_s
+\bar\omega)^2}.
\label{eq:general}\ee

\begin{figure}[t]
\centering
\vskip -0.3cm
\includegraphics[width=\colw]{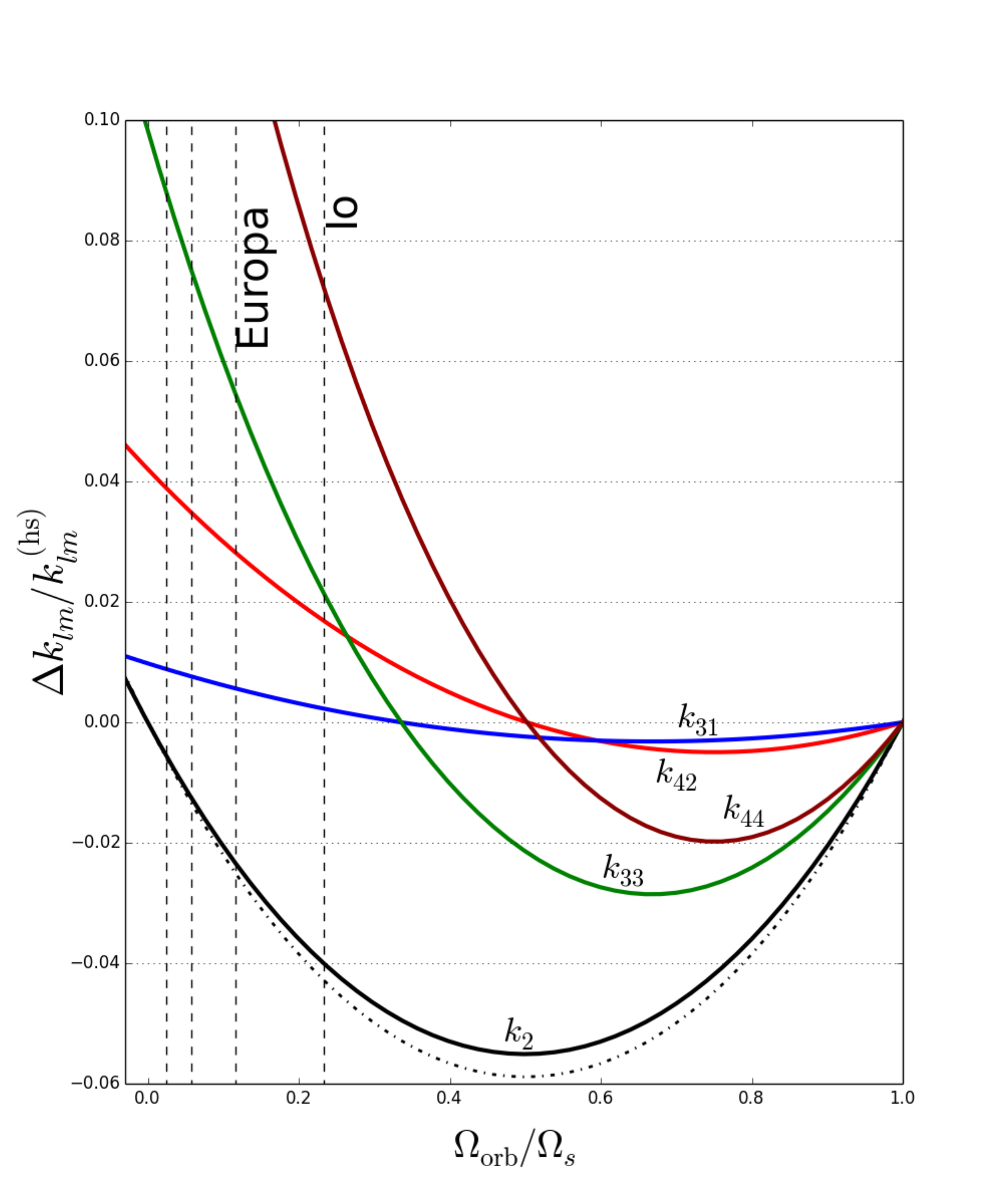}
\caption{\label{fig1}
  Dynamical correction $\Delta k_{lm}/k_{lm}^{\rm (hs)}=(k_{lm}-k_{lm}^{\rm (hs)})/k_{lm}^{\rm (hs)}$
  to Jupiter's tidal Love number as a funtion of the orbital frequency $\Omo$ of the perturbing satellite
  (in units of the spin frequency $\Oms$). All results (solid curves) are computed
  using the $n=1$ (isentropic) polytrope model, except that the dot-dashed curve is for $k_2=k_{22}$ computed
  using the $n=0.9$ polytrope model. The vertical dashed lines specify the orbital frequencies
  of Io, Europa, Ganymede and Callisto (from right to left). The $-4\%$ ``anomaly'' of $k_2$ observed by
  {\it Juno} can be explained by the planetary model with $n\simeq 1$.}
\end{figure}

For an incompressible planet model ($n=0$ polytrope), the $l=2$ mode (Kelvin mode) has
\be
\bar\omega_0={2\over\sqrt{5}},\quad
\bar Q_2=\left({3\over 2\pi}\right)^{1/2},\quad
C={1\over 2}.
\ee
Thus
\be
k_2\equiv k_{22}\simeq {3\over 2}\,\Bigl[1-{5\over 4}(\bar\Omega_s+\bar\omega)^2\Bigr]^{-1},
\ee
with $\bar\omega=2(\bar\Omega_{\rm orb}-\bar\Omega_s)$.

Giant planets are approximately described by a $n=1$ polytrope (corresponding to
$P\propto \rho^2$). Table 1
list the numerical values of $\omega_0,~Q_l$ and $C$ for several non-rotating
poltropic models (with different levels of stratification; see Section 5).
For $l=m=2$ tidal response (and $n=1$), $2C\simeq 1$, we have
\be
k_2\simeq 0.520\,\Bigl[1-0.664(\bar\Omega_s+\bar\omega)^2\Bigr]^{-1}.
\ee
Applying to the Jupiter-Io system:
Jupiter has $\bar\Omega_s=0.288$ [with spin period 9.925~hrs and
$2\pi (R^3/GM)^{1/2}=2.863$~hrs], Io has $\bar\Omega_{\rm orb}=0.0674$ (orbital period
1.769~days), so $\bar\omega=-0.441$. Thus the hydrostatic and dynamical $k_s$ values are
\be
k_2^{\rm (hs)}=0.550,\quad
k_2=0.528=0.960\,k_2^{\rm (hs)}
\ee
This explains the $4\%$ discrepancy between $k_2$ and $k_2^{(\rm hs)}$. Note that our static $k_2^{(\rm hs)}$
does not agree with the value (0.590) from Wahl et al.~(2020). This could arise for two reasons:
(i) The simple $n=1$ polytropic model does not precisely represent Jupiter's internal structure;
(ii) In deriving Eq.~(\ref{eq:general}), we have neglected order $\bar\Omega_s^2$ corrections to
the mode frequency and the tidal overlap\footnote{Both $\bar\omega_0^2$ and $\bar Q_l^2$
  in Eq.~(\ref{eq:general}) can have corrections of order $\bar\Omega_s^2$. These corrections
  do not affect $\Delta k_2/k_2^{\rm (hs)}$ to the leading order. Note that for $n=0$ (incompressible
  MacLaurin spheroid), the exact expressions for the mode frequency and tidal overlap coefficient
  are available [see Eq.~(3.4) and Eq.~(3.22) of Ho \& Lai (1999)].}.


Figure 1 shows the dynamical corrections
$\delta_{lm}\equiv \Delta k_{lm}/k_{lm}^{\rm (hs)}=(k_{lm}-k_{lm}^{\rm (hs)})/k_{lm}^{\rm (hs)}$
to Jupiter's tidal Love numbers as a function of the orbital frequency $\Omo$ of the perturbing satellite.
These results are obtained using the $n=1$ isentropic model, which gives the hydrostatic values
$k_{lm}^{\rm (hs)}\simeq 0.550,\,0.213,\,0.219,\,0.121,\,0.123$ for $(lm)=(22),\,(31),\,(33),\,(42),\,(44)$,
respectively. Although these hydrostatic values may not correspond to the ``true'' values for Jupiter
because of the simplicity of the polytrope model and the $\bar\Omega_s^2$ corrections (see above), 
the dynamical corrections $\delta_{lm}$ shown in Fig.~1 are robust.

Our results depicted in Fig.~1 can be compared to those of Idini \&
Stevenson (2021) obtained using more complicated calculations (see their Table 2).
Our $\delta_{22}$, $\delta_{33}$ and $\delta_{44}$ values (evaluated for the orbital
frequencies of Io, Europa, Ganymede and Callisto) agree reasonably
well with theirs, but our $\delta_{31}$, $\delta_{42}$ values are a
factor of a few smaller.

Finally, using Table 1, we can easily check that the contributions from p-modes
to $k_{\rm lm}$ are negligible.

\section{Inertial-Mode Contribution}

In addition to f-modes and p-modes, a rotating planet possesses a
spectrum of inertial modes supported by Coriolis force. 

For $n=1$ polytrope, the $m=2$ inertial modes have been computed 
by Xu \& Lai (2017) using a spectral code. The mode properties are
\be
\omega_+=0.556\Oms,~\varepsilon_+=0.28\Oms,~\bar Q_+=0.015\bar\Omega_s^2
\ee
for the prograde mode, and
\be
\omega_-=-1.10\Oms,~\varepsilon_-=-0.55\Oms,~\bar Q_-=0.010\bar\Omega_s^2
\ee
for the retrograde mode, where $Q_\pm$ is the tidal coupling coefficient $Q_{\alpha,22}$.
Since $\omega<0$, we can write the inertial mode contribution of $k_2$ as
\be
k_{2,\rm in}={2\pi\over 5}\left[{\bar Q_+^2\over \bar\varepsilon_+
    (\bar\omega_+ + |\bar\omega|)}+{\bar Q_-^2\over |\bar\varepsilon_-|
    (|\bar\omega_-| - |\bar\omega|)}\right].
\ee
Define $\hat\omega\equiv \omega/\Omega_s$ (and similarly $\hat\omega_\pm$ and
$\hat\varepsilon_\pm$) and $\hat Q_\pm \equiv \bar Q_\pm/\bar\Omega_s^2$, we have
\be
k_{2,\rm in}={2\pi\over 5}\,\bar\Omega_s^2 \left[{\hat Q_+^2\over \hat\varepsilon_+
    (\hat\omega_+ + |\hat\omega|)}+{\hat Q_-^2\over |\hat\varepsilon_-|
    (|\hat\omega_-| - |\hat\omega|)}\right].
\ee
For $n=1$ polytrope, this gives
\be
k_{2,\rm in}={2\pi\over 5}\,\bar\Omega_s^2 \times 10^{-4}\left({8.04\over 
    0.556 + |\hat\omega|} + {1.82\over 
    1.10 - |\hat\omega|}\right).
\ee
For Jupiter-Io system, $\hat\omega =2\Omo/\Oms -2 =-1.53$, it is clear that $k_{2,\rm in}\ll 1$.
In general, unless $|\hat\omega|$ happens to be very close to $|\hat\omega_-|$ (to within
$10^{-4}$), the contribution of the inertial modes to the Love number is negligible.

\section{Stable Stratification and G-Mode Contribution}

\begin{figure}[t]
\centering
\vskip -0.3cm
\includegraphics[width=\colw]{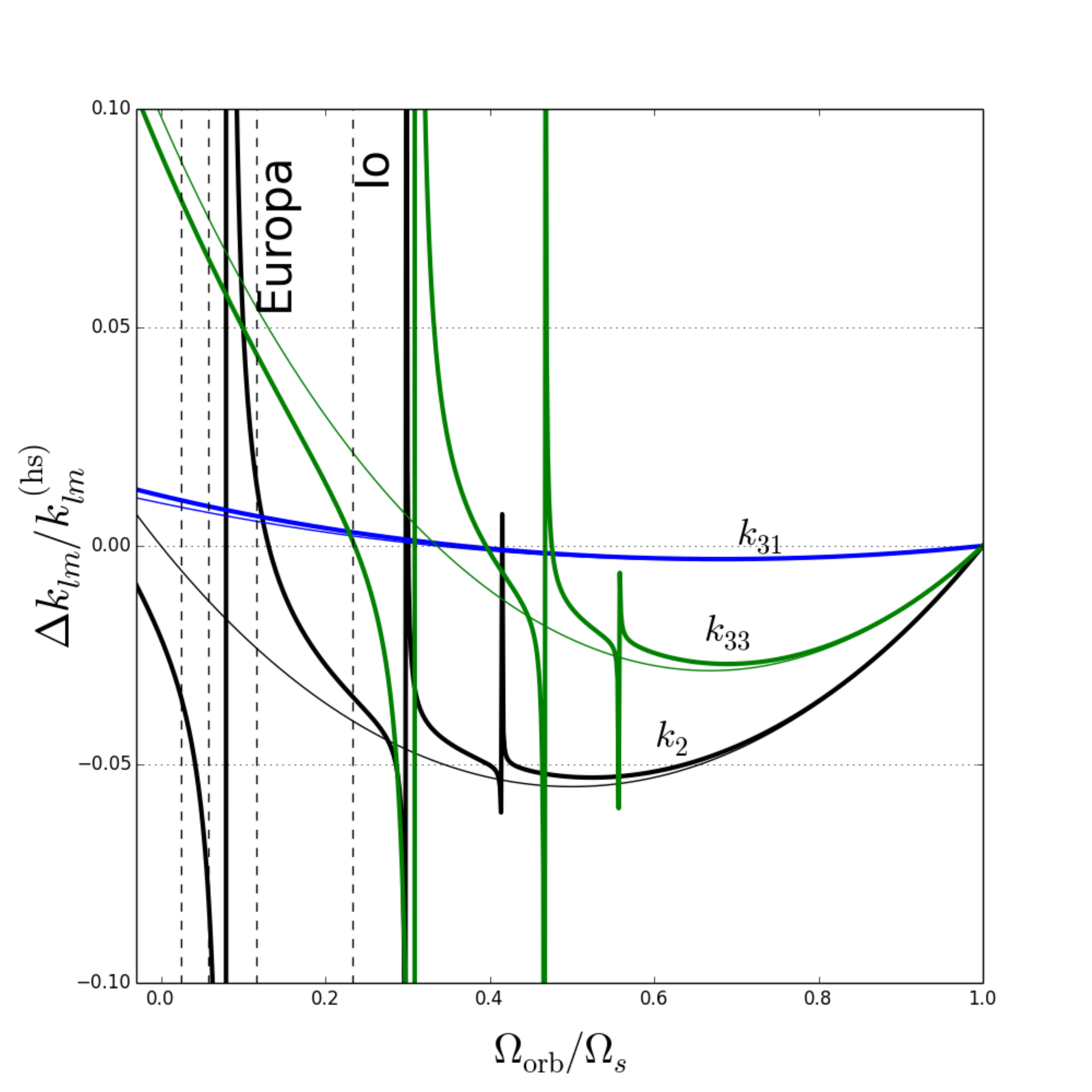}
\caption{\label{fig2}
  Same as Fig.~1, but for the $n=1$, $\Gamma_1=2.4$ planetary model, which possesses
  g-modes. The heavy solid curves include the contributions of f-modes and first three radial-order
  g-modes to $k_{lm}$, the light solid curves include only f-modes (for the $n=1$ isentropic model, as
  in Fig.~1).}
\end{figure}
\begin{figure}[t]
\centering
\vskip -0.3cm
\includegraphics[width=\colw]{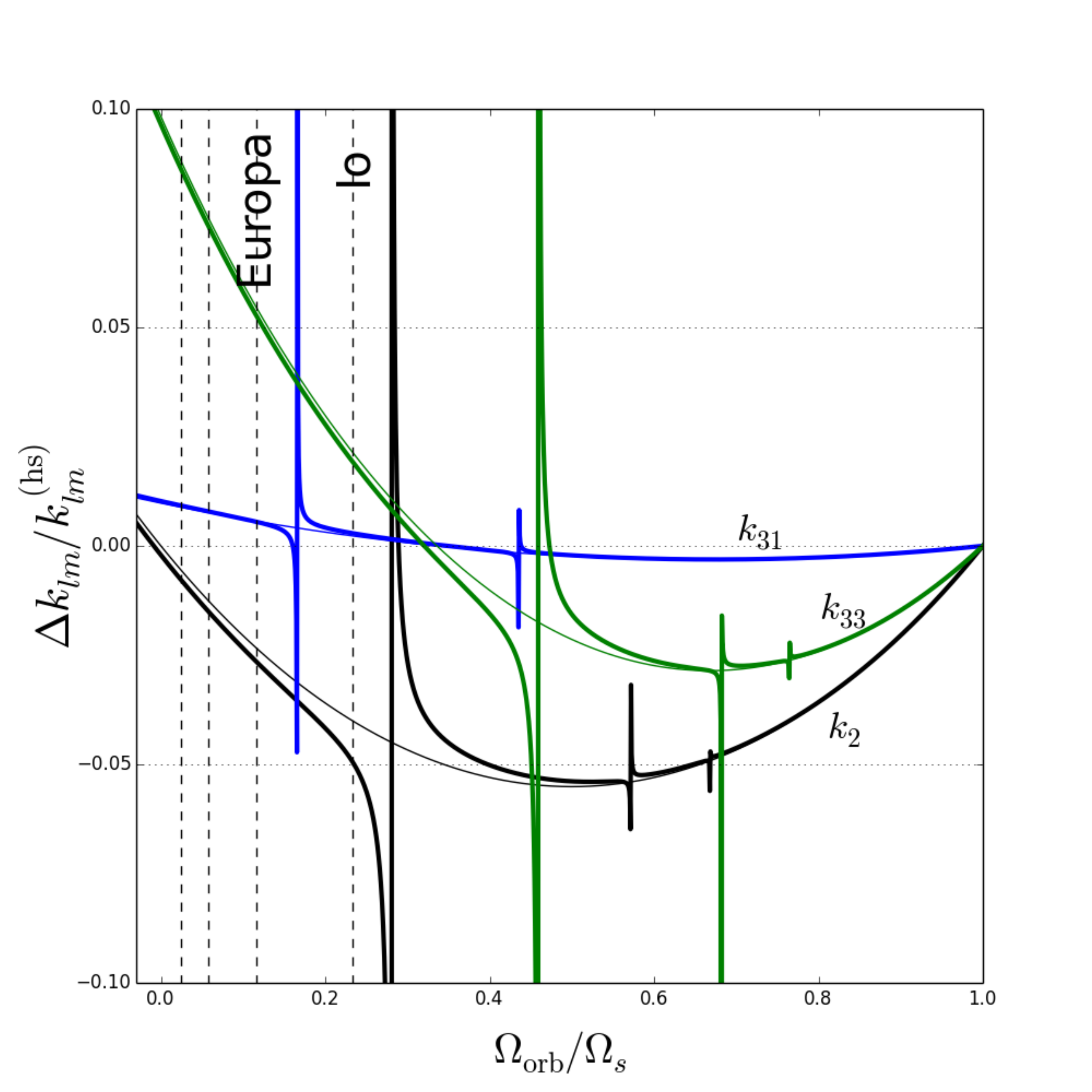}
\caption{\label{fig3}
  Same as Fig.~2, but for the $n=1$ model with stable stratification only in the
  region $r/R \in [0.5,0.7]$ (see Eq.~\ref{eq:gam1}).}
\end{figure}
\begin{figure}[t]
\centering
\vskip -1.cm
\includegraphics[width=\colw]{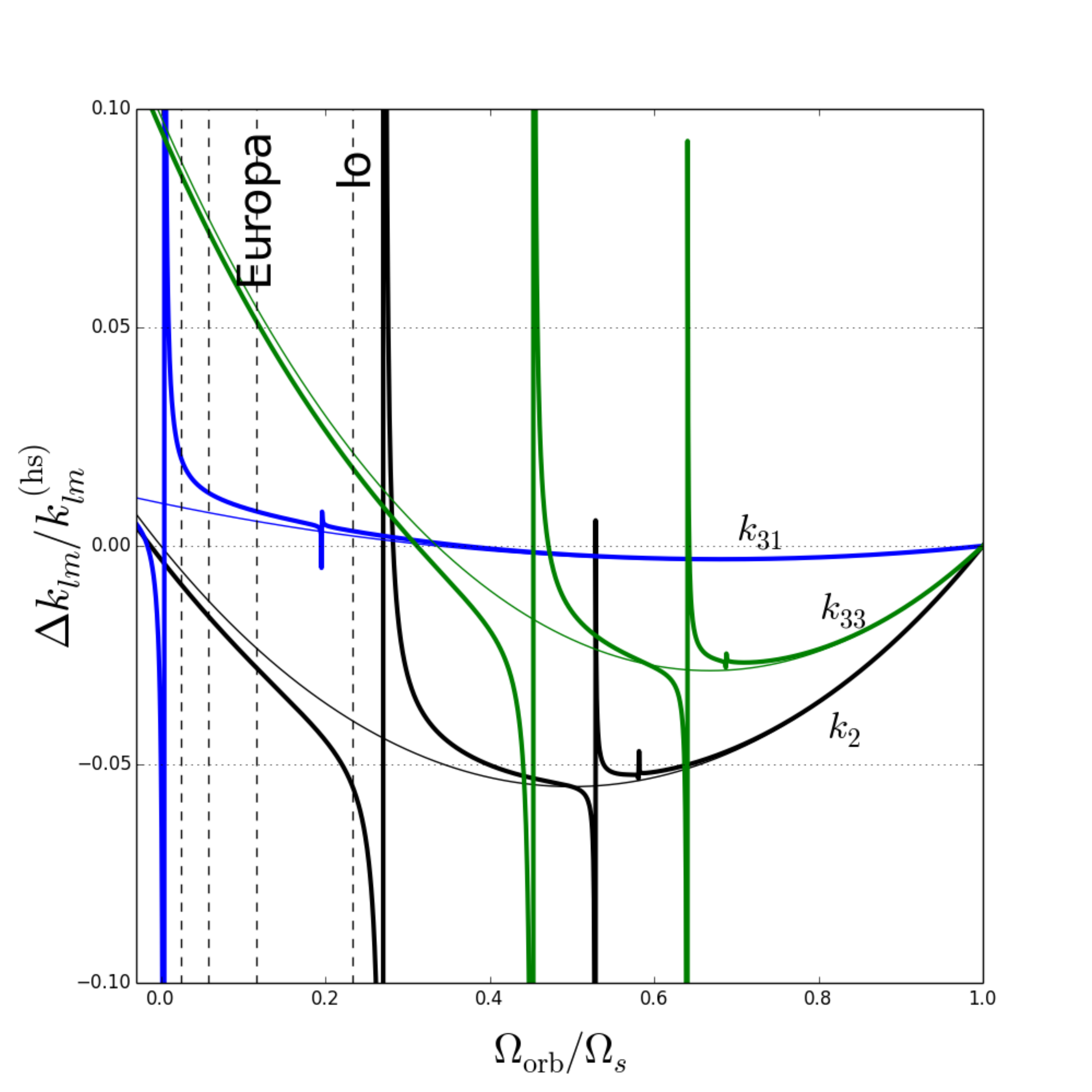}
\caption{\label{fig4}
  Same as Fig.~2, but for the $n=1$ model with stable stratification in two regions:
  $r/R \in [0.5,0.7]$ and $[0.85,0.93]$.}
\end{figure}

In Sections 3-4 we considered fully isentropic models for Jupiter, i.e., the adiabatic
index $\Gamma_1\equiv (\partial \ln P/\partial\ln\rho)_s$ equals the polytropic
index $\Gamma\equiv d\ln P/d\ln\rho=1+1/n$. In reality, some regions of the planet
may be stably stratified, with $\Gamma_1>\Gamma$. Indeed, the gravity measurement
by {\it Juno} and structural modeling suggest that Jupiter 
have a diluted core and a total heavy-element mass of 10-24 
Earth masses, with the heavy elements distributed within an extended region
covering nearly half of Jupiter’s radius (Wahl et al.~2017; Debras \& Chaberier 2019;
Stevenson 2020). The composition gradient outside the diluted core
would provide stable stratification, and the planet would then possess g-modes.
Another stable region may exist between (0.8-0.9)$R$ and $0.93R$ (Debras \& Chaberier 2019).

To explore of how g-modes influence the tidal love numbers, we consider three
simple planetary models, all having a $n=1$ density profile ($\Gamma=2$), but
with different adiabatic index profiles:
(i) $\Gamma_1=2.4$ throughout the planet (see Table 1);
(ii) $\Gamma_1=2.4$ only in the stable region $\bar r=r/R \in [0.5,0.7]$ (with a
transition width of 0.025) and $\Gamma_1=\Gamma=1+1/n$
otherwise, i.e.,
\be
\Gamma_1(r)=2+ {0.4 \over [1+e^{40(\bar r-0.7)}] [1+e^{-40(\bar r-0.5)}]}
\label{eq:gam1}\ee
(iii) $\Gamma_1=2.4$ only in two stable regions $\bar r=r/R \in [0.5,0.7]$ and
$[0.85,0.93]$ (with a transition width of 0.025) and $\Gamma_1=\Gamma=1+1/n$
otherwise (see Table 1).

For each model, we compute the f-modes and g-modes of a nonrotating
planet (see Table 1), and use Eqs.~(\ref{eq:eps})-(\ref{eq:omeg}) to
account for the effect of rotation on the modes.  We include only the
first three radial-order g-modes in our calculation of $k_{lm}$.  The
perturbative approach of the rotational effect is approximately valid
for these modes since $mC\Omega_s$ is less than the mode frequency
$|\omega_0|$.

Figures 2-4 show the results for the dynamical Love numbers based on the three models.
It is obvious that significant dynamical correction to the hydrostatic $k_{lm}^{\rm (hs)}$
occurs around the resonance, where $\omega_\alpha=\omega$.
The ``strength'' of each resonance is measured by the tidal overlap coefficient, and a large $|Q_l|$
value implies that the ``width'' of the resonant feature is larger (see Eq.~\ref{eq:general}).
For the $\Gamma_1=2.4$ model, the stratification is strong, the broad/strong resonance with the
g$_1$ mode can affect $k_{lm}$ associated with the Galilean moons (Fig.~2).
For the model with the stable stratified region restricted to $r/R\in [0.5,0.7]$ (Fig.~3), 
the resonance feature is much weaker/narrower, but still $\Delta k_2/k_2^{\rm (hs)}$ becomes $-5\%$ for Io.
When the model further includes the stratified region at $r/R\in [0.85,0.93]$ (Fig.~4), 
the resonance features shift and broaden, and $k_{31}$ becomes affected for the Galilean moons.
Obviously, these results are for illustrative purpose, but they indicate that resonance features
due to stable stratification in the planet's interior may influence the the measured
dynamical Love numbers.

Note that Figures 2-4 do not include contributions from high-order
g-modes. These modes (with mode frequencies comparable to $\Oms$)
become mixed with inertial modes (so-called ``inertial-gravity''
modes; see Xu \& Lai 2017) and cannot be treated using Eqs.~(\ref{eq:eps})-(\ref{eq:omeg}).
However, because of their small tidal overlap coefficients, they are unlikely
to be important contributors to $k_{lm}$ except for the coincidence of an extremely close
resonance.

\section{Conclusion}

We have derived a general equation (Eq.~\ref{eq:klm}) for computing
the dynamical Love number $k_{lm}$ of a rotating giant planet in
response to the tidal forcings from its satellites. In most situations,
the Love number is dominated by the tidal response of f-modes, and the
general expression reduces to Eq.~(\ref{eq:general}), which can be
easily evaluated using the mode properties of nonrotating planet
models (see Table 1). We show that the $4\%$ discrepancy between the measured 
$k_2$ of Jupiter and the theoretical hydrostatic value can be naturally 
explained by the dynamical response of Jupiter's f-modes to the tidal forcing from Io
-- the key is to include the rotational (Coriolis) effect in the tidal response
in a self-consistent way. We also show that the contributions of the inertial modes
to the Love number $k_2$ are negligible.

We have also explored the effect of stable stratification in Jupiter's
interior on the Love numbers. If sufficiently strong stratification
exists in a large region of the planet's interior, g-mode resonances
may influence the dynamical Love numbers associated with the tidal
forcing from the Galilean moons. Thus, precise measurements of
various $k_{lm}$ could provide constraints on the planet's
interior stratification.



\end{document}